\documentclass[a4]{article}

\usepackage{type1cm}        
\usepackage{makeidx}         
\usepackage{graphicx}        
\usepackage{multicol}        
\usepackage[bottom]{footmisc}
\usepackage[a4paper, margin={1.2in}]{geometry}

\usepackage{newtxtext}       %
\usepackage[varvw]{newtxmath}       
\usepackage{hyperref}

\makeindex             

\newcommand{\dee}{{\rm d}}
\newcommand{\R}{\mathbb{R}}

\newcommand{\norm}[1]{\left\vert#1\right\vert}
\newcommand{\Norm}[1]{\left\Vert#1\right\Vert}

\begin{document}

\title{Model-based reconstructions for quantitative imaging in photoacoustic tomography}
\author{Andreas Hauptmann\footnote{Research Unit of Mathematical Sciences, University of Oulu, P.O.Box 8000, 90014 Oulu, Finland. 
Department of Computer Science, University College London, WC1E 6BT, London, UK}\, \  and Tanja Tarvainen\footnote{Department of Technical Physics, University of Eastern Finland, P.O. Box 1627, 70211 Kuopio, Finland. 
Department of Computer Science, University College London, WC1E 6BT, London, UK}}
%
%
\maketitle


\abstract{The reconstruction task in photoacoustic tomography can vary a lot depending on measured targets, geometry, and especially the quantity we want to recover. Specifically, as the signal is generated due to the coupling of light and sound by the photoacoustic effect, we have the possibility to recover acoustic as well as optical tissue parameters. This  is referred to as quantitative imaging, i.e, correct recovery of physical parameters and not just a qualitative image. \\
In this chapter, we aim to give an overview on established reconstruction techniques in photoacoustic tomography. 
We start with modelling of the optical and acoustic phenomena, necessary for a reliable recovery of quantitative values. Furthermore, we give an overview of approaches for the tomographic reconstruction problem with an emphasis on the recovery of quantitative values, from direct and fast analytic approaches to computationally involved optimisation based techniques and recent data-driven approaches.}

\section{Introduction}
Photoacoustic tomography (PAT) offers the possibility to examine biological tissue on a micrometer scale with high contrast by coupling two physical phenomena: light and sound \cite{li2009,wang2016}. 
The obtained tomographic images carry important structural and quantitative information, which can be used to derive further biological and clinical markers, such as blood oxygen saturation \cite{beard2011,cox2012a}. 
For the correctness of such markers, it is important to have a robust and reliable reconstruction procedure available. 

While the underlying reconstruction problem is stable under ideal measurement setups, i.e., a full boundary coverage and fine sampling with a large bandwidth, the need for faster and more versatile imaging domains creates the need for more advanced reconstruction methods. 
In recent years, a multitude of different approaches have been introduced for the reconstruction problem in PAT, tackling specifically the acoustic or optical part, or both combined \cite{cox2012a,poudel2019survey,wang2023optical}. 
In this chapter, we will provide an introductory overview of existing reconstruction approaches, their applicability and possible limitations. We will put a special emphasis on quantitative reconstruction of correct physical parameters.

To understand the reconstruction task, we first need to define the problem. Mathematically, tomographic reconstructions are understood as an inverse problem, i.e., determining the cause from a set of 
measurements, under the knowledge of the measurement process, or the involved physics \cite{kaipio2005Book}. That is, we aim to recover an image $f$ from noise corrupted measurements $g$ under knowledge of the measurement operator, or forward operator, $A:f\mapsto g$. The measurement equation is then simply given by 
\begin{equation}\label{eqn:forwProb}
    Af + \varepsilon = g,
\end{equation}
where $\varepsilon$ denotes measurement noise, possibly including also mechanical/electronic inaccuracies. The \emph{inverse problem} is now to recover the image $f$ from measurements $g$ under knowledge of the relationship in \eqref{eqn:forwProb}. This already tells one important aspect of inverse problems, that we need to have a good understanding of the measurement process and the involved physics. For this purpose, we will first review the \emph{forward problem} of photoacoustic tomography in section \ref{sec:forwProb}, i.e., the physics of the measurement process.

In the following sections, we will first discuss the acoustic reconstruction problem, that is the recovery of a tomographic image representing the initial pressure from the measured acoustic signal. This will provide a comprehensive overview of reconstruction methods and different paradigms available. We will then discuss the specifics of reconstructions for the optical problem and the recovery of 
optical parameters.
We will conclude with a discussion of possible problems that one might encounter in the reconstruction process, such as uncertainties on measurement parameters and the need for computational efficiency. Finally, we give a short outlook on the promise of data-driven techniques to overcome some limitations.

\section{Forward modelling}\label{sec:forwProb}

In photoacoustic tomography,  a short (ns) pulse of near-infrared light is used to illuminate the region of tissue of interest. 
As light propagates within the tissue, it is absorbed by light absorbing molecules (chromophores). 
This generates localised increases in pressure, that propagates through the tissue as an acoustic wave and is detected by ultrasound sensors on the boundary or outside the imaged target.
The propagation of the acoustic wave occurs about five orders of magnitude slower than the light absorption, and thus the optical and acoustic parts of the problem can be decoupled and treated separately \cite{li2009}.

The \emph{forward problem} in quantitative PAT is to solve for the photoacoustic time-series on the ultrasound sensors when the optical and acoustic properties and the input light are given. 
Solving this problem consist of computing solutions to the optical and acoustic forward problems.

The process starts with the \emph{optical} forward problem, where we need to compute the absorbed optical energy density when the optical properties of the medium and the input light are given.
Let us consider modelling of light propagation and absorption at a single wavelength of light.
A widely accepted model for light transport in tissues is given by the radiative transfer equation (RTE) \cite{ishimaru78a}. 
The RTE is a one-speed approximation of the transport equation, and thus  it assumes that the energy of the particles does not change in collisions and that the refractive index is constant  in the medium.
Let $\Omega \subset \R ^d, \, d=2 \, \rm{or} \, 3$ denote the physical domain with boundary $\partial \Omega$ and  let $\hat{s} \in S^{d-1}$ (on the sphere) denote a unit vector in the direction of interest. 
In quantitative PAT,  the time-independent RTE is utilised
\begin{align}
  \label{eq:rte_cw}
  & \hat{s}
    \cdot \nabla \phi (r,\hat{s}) + (\mu _s + \mu _a) \phi(r,\hat{s}) 
    = \,\mu _s \int _{S^{d-1}}  \Theta (\hat{s} \cdot 
    \hat{s}') \phi(r,\hat{s}') \dee \hat{s}', \quad r
    \in \Omega, \\
  \label{eq:boundarycondition_rte}
  & \phi(r,{\hat s}) = \left \lbrace
    \begin{array}{cl} \phi _0(r,\hat{s}), & r \in \epsilon
      _j, \quad \quad  \hat{s} \cdot {\hat n} < 0, \\ 
      0, & r \in \partial \Omega \backslash \epsilon _j,
      \, \hat{s} \cdot {\hat n} < 0, \end{array} \right.
\end{align}
where \mbox{$\mu _s(r)$} and \mbox{$\mu _a(r)$}
are the scattering and absorption coefficients of the medium, respectively, \mbox{$\phi (r,\hat{s})$} is the radiance, 
\mbox{$\Theta (\hat{s} \cdot \hat{s}')$} is the scattering phase
function, \mbox{$\phi _0(r,\hat{s})$} is a source at a position \mbox{$\epsilon _j \subset \partial \Omega$}, and ${\hat n}$ is an outward unit normal  \cite{arridge99,tarvainen06a}.
The scattering phase function \mbox{$\Theta (\hat{s} \cdot \hat{s}')$} describes the probability that a photon with an initial direction ${\hat s}'$ will have a direction ${\hat s}$ after a scattering event.
In biological tissue, the most commonly used phase function is the Henyey-Greenstein scattering function \cite{henyey41} 
\begin{equation}
  \label{eq:Henyey-Greenstein}
  \Theta(\hat{s}\cdot\hat{s}') =  \left\{
  \begin{array}{ll}
  \frac{1}{2\pi} \frac{1 - g^2}{\left(1 + g^2 - 2g\hat{s} \cdot
    \hat{s}'\right)}, &  \quad d = 2 \\
  \frac{1}{4\pi} \frac{1 - g^2}{\left(1 + g^2 - 2g\hat{s} \cdot
    \hat{s}'\right)^{3/2}}, & \quad d = 3.
  \end{array} \right.
\end{equation}
where $g$ is an anisotropy parameter, \mbox{$-1 < g < 1$}. 

Due to the computational complexity of the RTE, various approximations are generally utilised in optical imaging. 
In a highly scattering medium, the RTE is approximated with the diffusion approximation (DA).
In the DA, the radiance is approximated by
\begin{equation}
  \label{eq:relat_rteandda1}
  \phi (r,{\hat s}) \approx \frac{1}{\norm{S^{d-1}}}\Phi (r) -
  \frac{d}{\norm{S^{d-1}}}\hat{s} \cdot  \left( \kappa \nabla \Phi (r)
    \right) 
\end{equation}
where \mbox{$\Phi(r)$} is the photon fluence 
\begin{equation}
  \label{eq:photondensity}
  \Phi(r)=\int _{S^{d-1}}\phi(r,\hat{s})\dee \hat{s}.
\end{equation}
Further, \mbox{$\kappa = \left( d(\mu _a + \mu _s') \right) ^{-1}$} 
is the diffusion coefficient where \mbox{$\mu _s' = (1-g_1)\mu _s$} is
the reduced scattering coefficient and $g_1$ is the mean of the
cosine of the scattering angle \cite{arridge99,tarvainen06a}, that  
in the case of the Henyey-Greenstein scattering function 
is $g_1=g$.
By inserting the approximation (\ref{eq:relat_rteandda1}) and similar
approximations written for the source term and phase function into
equation (\ref{eq:rte_cw}) and following the derivation in
\cite{arridge99,ishimaru78a,tarvainen06a}, the DA is obtained
\begin{align}
  \label{eq:da}
  & -\nabla \cdot \kappa \nabla \Phi (r) + \mu _a \Phi(r) = 0,  \qquad r
    \in \Omega, \\
  \label{eq:boundary_da_rob_withsource}
 & \Phi (r) + \frac{1}{2\gamma _d} \kappa  \frac{\partial \Phi (r)}{\partial
    \hat{n}} = \left \lbrace \begin{array}{ll} \frac{I
        _s}{\gamma_d}, & r\in
        \epsilon _i, \\ 
        0, & r \in \partial \Omega \setminus
        \epsilon _i, \end{array} \right. 
\end{align}
where $I_s$ is a diffuse boundary current and $\gamma _d$ is a dimension-dependent constant which takes
values $\gamma _2 = 1/\pi$ and $\gamma _3 = 1/4$ \cite{tarvainen06a}. 
The DA is a valid approximation when the radiance is almost a uniform distribution, i.e. in a scattering dominated medium  further than a few scattering lengths from the light source \cite{ishimaru78a}.
In PAT, however, imaging depth can be small compared to the average scattering length, and thus the DA is not always a valid approximation.

Since light absorption and pressure increase occur significantly faster compared to acoustic wave propagation, they can be modelled instantaneous.  
The absorbed optical energy density $H(r)$ created by light absorption can be solved from photon fluence as
\begin{equation}
  \label{eq:absorbed_energy}
  H(r) = \mu_a(r) \Phi(r).
\end{equation}
Further, generated pressure can be approximated as an initial pressure $p_0$ that is connected to the absorbed optical energy density through the photoacoustic efficiency that can be identified with the Gr\"uneisen parameter $G$ for an absorbing fluid \cite{cox2012a}
\begin{equation}
  \label{eq:initial_pressure}
  p_0(r) = p(r,t=0) = G H(r).
\end{equation}
 
The following \emph{acoustic} forward problem in quantitative PAT is to solve for the photoacoustic time-series $p(r,t)$ at the sensors.
Time evolution of the photoacoustic wave can be modelled using the equations of linear acoustics \cite{cox2005,treeby2010kWave}.
For soft biological tissues, it is generally assumed that the medium is isotropic and quiescent and that shear waves can be neglected.
Then, the propagation of initial pressure can be described as an initial value problem in acoustics using an acoustic wave equation
\begin{align}
    \label{eq:wave_equation_pat}
   (\partial_{tt} - v^2 \Delta)p(r,t) &= 0, \\
  \label{eq:wave_initialcondition1}
   p(r,t=0) &= p_0(r), \\
  \label{eq:wave_initialcondition2}
   \partial_t p(r,t=0) &= 0.
\end{align}
The spatial distribution of the speed of sound $v$ depends on the target and is usually not known \emph{a priori}. 
Still, it is often assumed to be constant within the medium. 
The time-series $p(r,t)$ on the boundary of the domain now constitutes the measured signal in photoacoustic tomography, from which we will start the reconstruction process.

\section{Reconstruction of the photoacoustic image}
The reconstruction task in PAT follows the reverse order of the forward model, that is we are dealing with an inverse problem. Specifically, the forward problem starts with the optical illumination, followed by the generation of the acoustic signal through the photoacoustic effect. The inverse problem consequently starts with the measured  pressure wave on the sensor, from which a we can recover the initial pressure $p_0$, only after which we can solve the optical problem. Here, the first acoustic problem is linear, whereas the second optical problem is nonlinear. 

In the following we will first discuss the reconstruction task for the acoustic problem in more detail. The measurement is given by the pressure wave $p(r,t)$ on the boundary, i.e., $r\in \partial\Omega$ and a finite time interval $[0,T]$. Usually, the measurement is restricted to only a part of the boundary and we may only collect sub-sampled data. Additionally, the detector will have an angle-dependent frequency response which needs to be taken into account. For simplicity we assume here that this is encoded in the forward operator $A$ that maps the initial pressure to the measured time-series and we can simply write the measurement equation as
\begin{align}
\label{eq:measurement_equation}
    Af + \varepsilon = g, \text{ where  } &f(r)=p_0(r), \, r  \in \Omega, \\ &g(r,t)=p(r,t), \, r \in \partial\Omega,\, t\in[0,T], \nonumber
\end{align}
where $\varepsilon$ denotes additional measurement noise.
Below, we give a short overview of different inverse problem solution methodologies applied in PAT. For a more detailed review on reconstruction algorithms, we refer to a recent review \cite{poudel2019survey}.

\begin{figure}[tb!]
\centering
\includegraphics[width=0.7\textwidth]{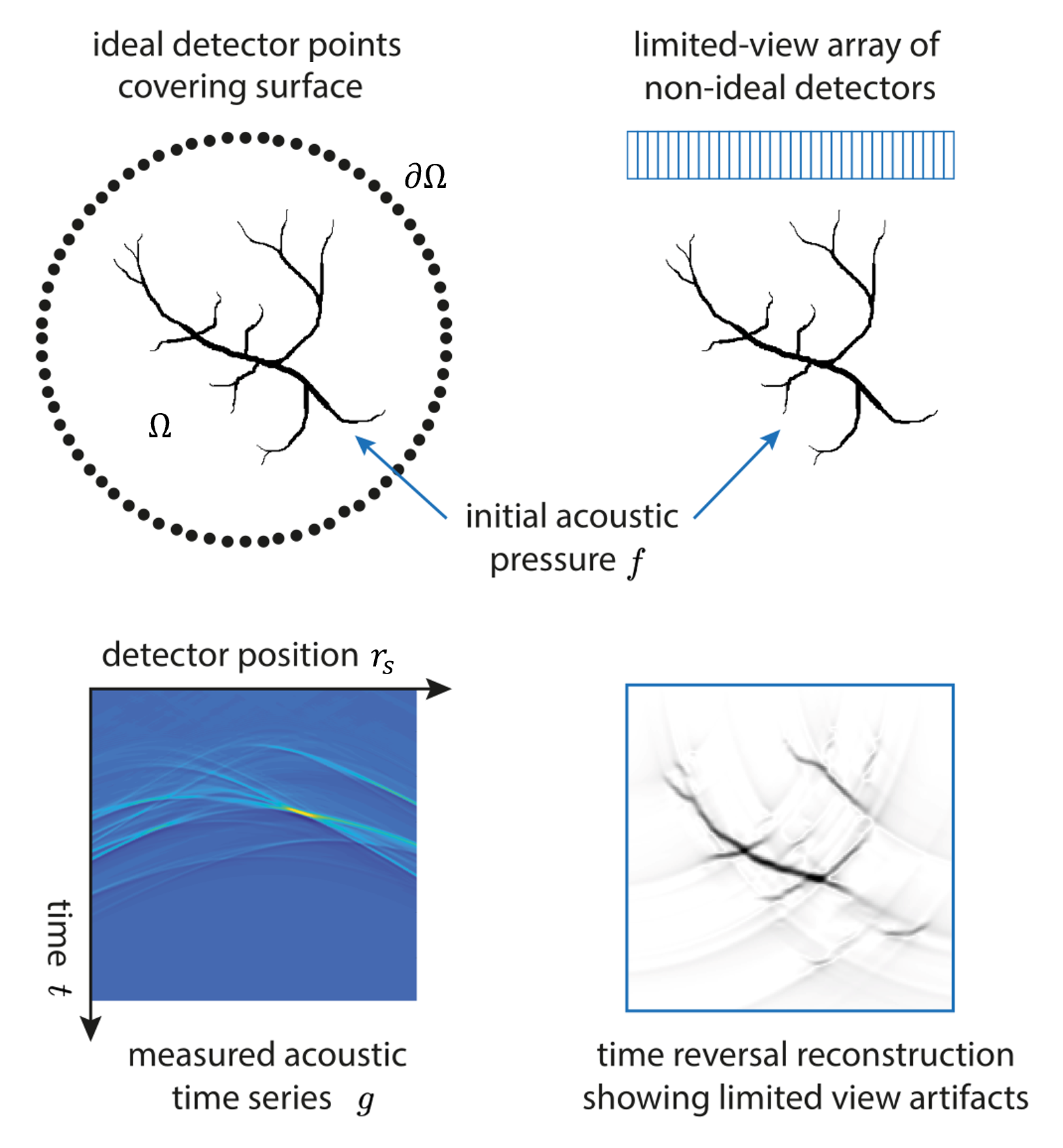}
\caption{\label{fig:measurement_setup} (Top left) Ideal PAT measurement setup in 2D with point-like omnidirectional detectors covering the
surface $\partial\Omega$ surrounding the initial pressure distribution $f$. (Top Right) A typical measurement setup using a finite-sized linear array
of detectors. (Bottom left) The acoustic time series measured by the linear array. (Bottom right) A PAT image reconstructed from these time series using the classical
time reversal approach showing arc-like artifacts due to the limited view detection. (Figure adapted from \cite{hauptmann2020deep})}
\end{figure}

\subsection{Direct reconstructions}
The acoustic inverse problem is well-posed, when full measurement data exists \cite{Xu2004}. That is, the measurement domain surrounds the whole target and the time-series is measured long enough, i.e., for all times where the pressure wave is non-zero. This is clearly easier to achieve in a two-dimensional setting than in 3D, where we often are limited to only a part of the boundary. This leads to what is known as a limited-view setting, see Fig. \ref{fig:measurement_setup} for an illustration.

In case there is sufficiently large part of the boundary covered, we can obtain good reconstructions by \emph{backprojection}-type algorithms. Here, the data $g$ is projected back along a set of spherical shells of radius $t = |r - r_s|/v$ centered on the detector points $r_s$. We can define the general class of backprojection operators 
 $A^{\#}$ that map the measured data $g(r_s,t)$ to the reconstructions $f(r)$ by projecting the data back on these shells by the mapping $t \rightarrow |r - r_s|/v$ followed by summation over all detector points $r_s$. Together with a function $h(r_s,t)$ that is either given by the measurement data or some transformation of it, depending on the measurement setup. The general backprojection operator is then given by
\begin{align}\label{eqn:backprojectionIntegral}
(A^{\#} g)(x) = \int_{\partial\Omega} \left[h(r_s,t)\right]_{t = |r - r_s|/v} \, ds(r_s) 
\end{align}
where $d{s}$ is an area element on the measurement surface $\partial \Omega$. 
Clearly, the recovery of quantitative values, in contrast to qualitative reconstructions, depends on the filter function $h(r_s,t)$. 
For instance, in case of the adjoint operator $A^*$ \cite{arridge2016adjoint} the backprojection takes the form
\begin{align}
(A^* g)(x) = \int_{\partial\Omega} \left[\frac{1}{4\pi |r-r_s|}\frac{\partial g}{\partial t}(r_s, t) \right]_{t = |r - r_s|/v} ds(r_s), 
\end{align}
which primarily provides a qualitative reconstruction with slightly over estimated contrast.

In case of the well-known `universal backprojection' algorithm \cite{xu2005universal}, we obtain exact reconstructions with correct quantitative values for spherical and cylindrical \cite{Finch:2004msph,kunyansky2007explicit} or even planar measurement surfaces that enclose the target.


\subsubsection{Time reversal}
The wave equation can be stably solved backwards in time and hence one can obtain a physically intuitive approach to reconstruction by so-called \emph{time reversal} \cite{hristova2008reconstruction,xu2004time}, but one needs to be careful of the boundary conditions. 
Given the measurement surface $\partial \Omega$ surrounding the measurement domain in which the target is supported $f\subset \Omega$. In the forward model, the photoacoustically-generated waves are propagating outwards and are measured on the surface $\partial\Omega$ until after suitably long time $T$ the acoustic field in $\Omega$ is zero. 

In the time reversal process, we assume that measured pressure $g(r_s,t)$ is now produced on $\partial\Omega$ in time-reversed order, starting with $g(x_s,T)$.
Thus, the acoustic field in $\Omega$ created by the \emph{in-going} waves reproduces the out-going wavefield exactly, but backwards in time. Consequently, the acoustic field at $t=0$ will reproduce the initial acoustic pressure $f(r)=p_0(r)$.
Numerically, time reversal image reconstruction solves the following time-varying boundary value problem for the time-reversed field $p_-(r,t_-)$, from time $t_- = 0$ to $T$,
\begin{align}
   (\partial_{tt} - v^2 \Delta)p_-(r,t_-) &= 0, \\
   p_-(r_s,t_-) &= g(r_s,T - t_-), \\
   p_- = \partial_t p_-(r,0) &= 0.
\end{align}
We then obtain at final time $T$ the reconstruction $p_-(r,T) = f(r)$ for $r\in\Omega$. A reconstruction by time reversal is shown in Fig. \ref{fig:measurement_setup} for a limited-view geometry in 2D.

\subsection{Variational methods}
In limited-view or when only sparsely sampled data is available, direct reconstructions often do not perform sufficiently well, qualitatively as well as quantitatively. In such scenarios it is advised to utilise so-called variational methods \cite{arridge2016accelerated,scherzer2009variational}. That is, the reconstruction $f^*$ is sought as minimiser of a cost functional that measures data-fit together with a regulariser,
\begin{equation}\label{eqn:VariationalForm}
f^* = \arg\min_f \|Af-g\|_2^2 + \alpha \Psi(f). 
\end{equation}
Here, the data-fidelity term $ \|Af-g\|_2^2$ naturally ensures that the reconstructed image fits the measured data and hence $f^*$ also provides correct quantitative values. The regulariser $\Psi(f)$ has a two-fold purpose, first of all it ensures well-posedness of the reconstruction task and stabilises the optimisation, additionally it allows to model prior knowledge on the target structures by penalising undesired features. Typical regulariser are, for example, Tikhonov regularisation $\Psi(f)=\|f\|_2^2$ that supports smoothness of the reconstruction, the total variation penalty $\Psi(f)=\|\nabla f\|_1$, which enforces reconstructions to be piece-wise constant and hence is highly effective in noise suppression \cite{arridge2016accelerated,bergounioux2014optimal,wang2012investigation}, higher order variants \cite{boink2018framework}, as well as generally regulariser based on the 1-norm promoting sparsity \cite{haltmeier2016compressed}. 
The parameter $\alpha>0$ in \eqref{eqn:VariationalForm} balances both terms and allows one to fine tune the desired reconstructions: a small parameter allows for emphasis on the data-fit (advised under small noise), whereas a large parameter enforces stronger regularity. In other words, we choose how much we trust the measured data.

Reconstructions are then computed by optimisation techniques that minimise \eqref{eqn:VariationalForm} iteratively. If both data-fidelity and regulariser are differentiable, gradient gradient descent methods can be utilised. Whereas the data-fidelity using the 2-norm is differentiable, popular choices for the regulariser are based on sparsity and are often based on the non-differentiable 1-norm. Thus, one often uses techniques from convex analysis\footnote{Note, that the problem \eqref{eqn:VariationalForm} \cite{Rockafellar1970} is convex if $A$ is linear and the regulariser $\Psi(f)$ is convex.} and optimisation to compute minimiser \cite{benning2018modern}. One such option are proximal gradient descent schemes. Here, we start with some initialisation $f_0=A^{\#}g$, we then iteratively take the gradient with respect to the differentiable data-fidelity, given by $\nabla_f\|Af-g\|_2^2=2A^*(Af-g)$, followed by the application of a proximal operator that incorporates the regularisation for convex $\Psi(f)$. The iterations are then given by   
\begin{equation}\label{eqn:prox_gradDesc}
    f_{k+1} = \mathrm{prox}_{\Psi,\alpha\lambda_k} \left(f_k - \lambda_k A^*(Af_k-g)\right),
\end{equation}
where the proximal operator projects solutions to the admissible space of the regulariser by solving another simpler optimisation problem
\begin{equation}\label{eqn:prox_op}
    \mathrm{prox}_{\Psi,\alpha\lambda_k}(f) = \arg\min_h \left\{\Psi(h) + \frac{1}{2\alpha\lambda_k} \|h-f\|_2^2 \right\}.
\end{equation}
The advantage of the proximal operator are two-fold as well, first for some choices of $\Psi(f)$ there is a closed form solution, such as the 1-norm $\Psi(f)=\|f\|_1$. Second, the problem \eqref{eqn:prox_op} does not involve the forward operator $A$ and hence is a denoising problem for which many efficient algorithms exist, this is the case for total variation.

\begin{figure}[t!]
\centering
\includegraphics[width=0.8\textwidth]{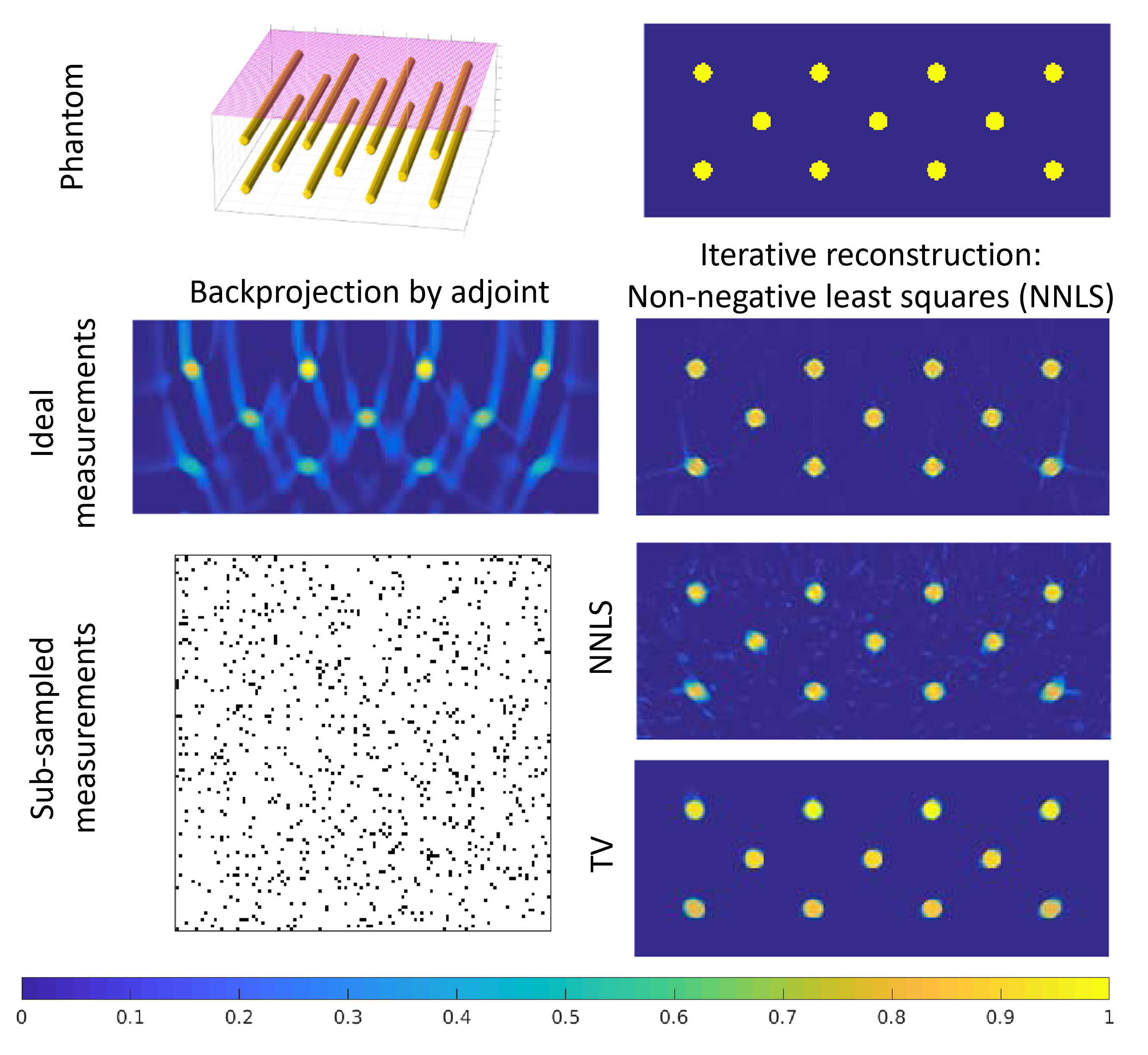}
\caption{\label{fig:recons} Reconstructions in 3D from fully-sampled and sub-sampled limited-view measurements. (Top Row) The phantom in 3D and cross section: the measurement sensor is marked in red on the top. (Middle row) Reconstruction from ideal fully-sampled measurements: backprojection by the adjoint on the left and an iterative reconstruction using only non-negativity as regularizer on the right. (Bottom row) Reconstructions from sub-sampled data: The sub-sampling pattern is shown on the left and two reconstructions with non-negativity (NNLS) and total variation (TV) on the right. (Figure adapted from \cite{hauptmann2018model}) }
\end{figure}

An illustration of reconstructions using the variational approach is shown in Fig. \ref{fig:recons}. The backprojection clearly leaves visible limited-view artefacts and loss of contrast in the deeper areas. Whereas, iterative reconstructions can effectively remove those by iteratively enforcing data consistency. When fully sampled data is available, just enforcing non-negativity as regulariser is sufficient, for sub-sampled measurements a stronger regulariser is needed, such as total variation enforcing piece-wise constant targets, effectively eliminating the residual noise.

\subsection{Bayesian methods}

In a Bayesian approach, the inverse problem is solved in the framework of statistical inference \cite{kaipio2005Book}. 
All parameters are considered as random variables, and information about these parameters is expressed by their probability distributions.
In the inverse problem, the idea is to obtain information about the parameters of primary interest based on the measurements, the model, and the prior information about the parameters.
The solution of the inverse problem is then given by the posterior probability distribution $\pi(f|g)$, that according to Bayes' theorem can be presented as a conditional probability density function of the form
\begin{equation}
    \label{eq:Bayes_theorem}
    \pi (f|g) \propto   \pi (g|f) \pi (f),
\end{equation}
where $\pi(g|f)$ is the likelihood density and $\pi(f)$ is the prior density.
Considering the observation model with additive noise \eqref{eq:measurement_equation}, the likelihood probability distribution can be written as 
\begin{equation}
    \label{eq:likelihood_density}
    \pi (g|f) = \pi_{\varepsilon} (g-Af)
\end{equation}
where $\pi_{\varepsilon}(\cdot)$ is the probability density of the noise $\varepsilon$.
Let us model noise $\varepsilon$ and unknown $f$ as Gaussian distributed, i.e. $\varepsilon \sim \mathcal{N}(\eta_{\varepsilon},\Gamma_{\varepsilon})$ and $f \sim \mathcal{N}(\eta_f,\Gamma_f)$ where $\eta_{\varepsilon}$ and $\Gamma_{\varepsilon}$ are the mean and covariance of the noise and $\eta_f$ and $\Gamma_f$ are the mean and covariance of the prior for the unknown $f$.
In this case, the posterior probability density can be written as
\begin{equation}
    \label{eq:posterior_density}
    \pi (f|g) \propto \exp \left\lbrace -\frac{1}{2} \Norm{L_{\varepsilon}( g-Af-\eta_{\varepsilon} )}^2 - \frac{1}{2} \Norm{L_f ( f-\eta_f)}^2 \right\rbrace,
\end{equation}
where $L_{\varepsilon}^{\rm T}L_{\varepsilon}=\Gamma_{\varepsilon} ^{-1}$ and $L_f^{\rm T}L_f = \Gamma_f^{-1}$ are the square roots such as the Cholesky decompositions of the inverse covariance matrices of the noise and prior, respectively.

In the case of a linear observation model and Gaussian noise and prior, the posterior density is also a Gaussian distribution \cite{kaipio2005Book,tick2016image}
\begin{equation}
    f|g \sim \mathcal{N}(\eta_{f|g}\Gamma_{f|g}), \nonumber
\end{equation}
where the mean $\eta_{f|g}$ and covariance $\Gamma_{f|g}$ can be written in the form
\begin{align}
    \label{eq:posterior_mean}
    \eta_{f|g}  = K^{-1}b \text{ and }
    \Gamma_{f|g}  = K^{-1},
\end{align}
with
\begin{align}
    \label{eq:posterior_term1}
    K & = A^{\rm T}\Gamma_{\varepsilon}^{-1}A + \Gamma_f^{-1}, \\
    \label{eq:posterior_term2}
    b & = A^{\rm T} \Gamma_{\varepsilon}^{-1}(g-\eta_{\varepsilon})+\Gamma_f^{-1}\eta_f.
\end{align}
In case solving the posterior distribution directly using Eqs. \eqref{eq:posterior_mean}--\eqref{eq:posterior_term2} is computationally too expensive, it can be evaluated using a matrix free approach, for example \cite{tick2018three}. The advantage of a Bayesian approach lies in the additional information on mean and variances of the computed solutions, which allows for uncertainty quantification giving estimates on the correctness of solutions. 

\subsection{Data-driven reconstruction methods}
In recent years, data-driven approaches have gained considerable interest due to their ability to learn data-specific representations. That means, instead of designing hand-crafted analytic priors (in the variational setting) or statistical priors (in the Bayesian setting), one can now utilise large quantities of representative data. In most cases, these are given by ideal reconstructions, but here also lies the primary limitation in photoacoustic tomography, since large quantities of ideal or ground-truth reconstructions are not easy to collect, especially in three-dimensions. Additionally, at the current stage there are limited guarantees available on the correctness of the reconstructions apart from empirical evaluations. In the following we give a short summary of the underlying paradigm of learned reconstruction methods in the context of photoacoustic tomography, for a complete review we refer to \cite{grohl2021deep,hauptmann2020deep,yang2021review}.

In general terms, the aim of learned image reconstruction is to design a parameterised reconstruction operator $R_\theta$ with learnable parameters $\theta$, such that
\begin{equation}\label{eqn:reconOp}
    R_\theta(g)\approx f.
\end{equation}
Clearly, there are many ways to define such a learned reconstruction operator, as there are many classical reconstruction approaches available as well. We can roughly group approaches into three classes:
\begin{itemize}
    \item[i.] The fully learned approach: Here, $R_\theta$ is purely given as a neural network $G_\theta: g \mapsto f$ that directly transform the measurements into reconstructions. See, for examples \cite{shang2021two,waibel2018reconstruction}.
    \item[ii.] The two-step approach: First a classical reconstruction is obtained by, e.g., a backprojection $A^{\#}: g\mapsto \widetilde{f} $, which is usually corrupted by noise, undersampling and possible limited-view artefacts. A neural network $G_\theta:\widetilde{f}\mapsto f$ is then trained to remove these artefacts \cite{antholzer2019deep}. 
    \item[iii.]  Learned iterative (model-based): Here, neural networks and the forward/adjoint operator are intertwined in an iterative way \cite{hauptmann2018model}. That is, the reconstruction is updated repeatedly motivated by the variational formulation. This can be achieved, for instance, by replacing the proximal operator in \eqref{eqn:prox_gradDesc} with a neural network, such that $G_\theta:f_k\mapsto f_{k+1}$.
\end{itemize}
Generally speaking, while the model information increases from i. to iii. the complexity of the learning task decreases, i.e., the network needs to learn an easier operation. This decrease in complexity does also reduce the quantity of necessary data to train the network, but comes with an increase of computational complexity due to the involvement of $A$. In practice, it is important to choose the most suitable approach for ones needs.

When aiming for correct and reliable quantitative reconstructions with learning based approaches, there are unfortunately only limited guarantees on correctness. At this stage, the majority of approaches does rely on empirical validation rather than theoretical guarantees. Nevertheless, current theoretical research concentrates on providing reconstruction guarantees. This can be primarily achieved by restricting the networks $G_\theta$, but usually also reduces expressivity of the network and hence results in a reduction of performance, see \cite{mukherjee2023learned} for a recent survey on the subject of provably convergent learned reconstructions.

\section{The inverse problem of quantitative photoacoustic tomography}

As the second step of quantitative photoacoustic imaging, optical parameters are estimated from the photoacoustic images obtained as the solution of the acoustic inverse problem.
This second inverse problem includes a nonlinear forward model describing light propagation in tissues.
Most of the approaches for this optical inverse problem have been based on using the DA \eqref{eq:da}--\eqref{eq:boundary_da_rob_withsource} as light transport model.
However,  the RTE \eqref{eq:rte_cw}--\eqref{eq:boundarycondition_rte} \cite{haltmeier2015single, mamonov2014, pulkkinen2015quantitative,saratoon2013,tarvainen2012} and Monte Carlo method for light transport \cite{buchmann2020quantitative,buchmann2019three, hanninen2022adaptive, hochuli2016quantitative,leino2020perturbation, macdonald2020efficient} have also been utilised.
In addition to modelling light propagation and absorption, photoacoustic efficiency identified by the Gr{\"u}neisen parameter, Eq. \eqref{eq:initial_pressure}, needs to be taken into account.

Estimation of more than one optical parameter is a non-unique problem if only one light illumination is used.  
To overcome this, one approach has been to assume the scattering as known  \cite{buchmann2020quantitative,  cox2006two, jetzfellner2009performance,ripoll2005quantitative}.  
However, in practical applications scattering is typically not known. 
It has been shown that the non-uniqueness can be overcome by using multiple optical illuminations \cite{bal2011a}.  
Also combining photoacoustic with diffuse optical tomography (DOT) data have been used to ease the non-uniqueness \cite{li2013impact, nykanen2017,ren2013hybrid}.

\subsection{Optical inverse problem}

The aim in quantitative PAT is to  estimate the concentrations of chromophores. 
Generally, the absorption coefficient can be expressed as a linear combination of the chromphore concentrations
\begin{equation}
    \label{eq:absorption_chromophores}
    \mu_a(\lambda) = \sum_{k=1}^K \alpha_k(\lambda) C_k,
\end{equation}
where $\alpha_k$ and $C_k$ are the absorption coefficient and concentration of the $k$th chromophore, and $\lambda$ is the wavelength of the light.
Estimation of the concentrations of chromophores can be achieved either by directly estimating the concentrations from photoacoustic time-series obtained using different wavelengths of light  \cite{bal2012,cox2009,laufer2010,mamonov2014,pulkkinen2014,razansky2009,razansky2012} or by
first recovering the absorption coefficients at different wavelengths
and then calculating the concentrations utilising \eqref{eq:absorption_chromophores} for the absorption spectrum
\cite{bal2012,cox2012a,cox2009,pulkkinen2014}. 
Sometimes, researchers are interested in secondary quantities that can be computed from the optical parameters \cite{bigio2016book}. 
One such quantity of high clinical interest is blood oxygen saturation ($sO_2$) \cite{Li2018reviewSO2}, which is related to the concentrations of the two endogenous chromophores, oxy- and deoxy-hemoglobin.
Different approaches for solving the optical inverse problem have been proposed.
Here, we give a short summary of those approaches. 
For more information, see e.g. the recent review \cite{wang2023optical} and the references therein.

Perhaps the most simple approach for estimating the optical absorption in quantitative PAT has been to assume the photon fluence as known or solvable using some predefined optical parameter values, and then solve the absorption utilising so-called fluence correction methods \cite{zhou2020evaluation}.
The drawback of these approaches is that both absorption and scattering affect photon fluence, and thus approximating these as preassigned parameters creates modelling errors. 
Alternatively, linearised models leading to simple solvers for estimating absorption \cite{cox2006two,zemp2010} or absorption and scattering simultaneously \cite{shao2011} have been utilised.

In most studies on the optical inverse problem, simultaneous estimation of absorption and either scattering or diffusion have been considered.
The inverse problem has been approached, for example, using direct methods \cite{bal2011a,bal2012,mamonov2014}.
Similarly as in the acoustic problem, a commonly utilised approach has been to estimate the optical parameters $x$ by formulating the problem as a regularised minimisation problem 
\begin{equation}
    \label{eq:optical_minimisation}
    x^{\ast} = \arg\min_x \Norm{B(x)-y}_{2}^2 + \beta \Psi (x) 
\end{equation}
where $B$ denotes the optical forward operator, $y$ is the data of the optical inverse problem, $\beta$ is the regularisation parameter and $\Psi$ is a regularising penalty functional.
Typically, Tikhonov regularisation \cite{saratoon2013} supporting smoothness of the solution, total variation \cite{gao2010bregman, hannukainen2016efficient,tarvainen2012} promoting piece-wise constant parameters, and $\ell_1$-norm regularisation \cite{zhang2014forward} promoting sparsity in the solution have been utilised. 
The minimisation problem \eqref{eq:optical_minimisation} has been solved using methods of computational optimisation.
Alternatively, a Bayesian approach leading to a \emph{maximum a posteriori} (MAP) estimation problem 
has been taken in \cite{hanninen2022adaptive,leino2020perturbation, nykanen2017,pulkkinen2014,tarvainen2013bayesian} enabling inclusion of quantitative prior models and modelling of noise and uncertainties.
Furthermore, examples of alternative approaches to the optical inverse problem include  \cite{alberti2017disjoint,rosenthal2009quantitative} where the absorption and the photon fluence were extracted using a sparse signal representation.
Further, in \cite{naetar2014quantitative}  boundaries between piece-wise constant optical parameters were estimated, and in \cite{malone2015reconstruction} a combined reconstruction-classification approach was proposed.

\begin{figure}[tb!]
\centering
\includegraphics[width=0.9\textwidth]{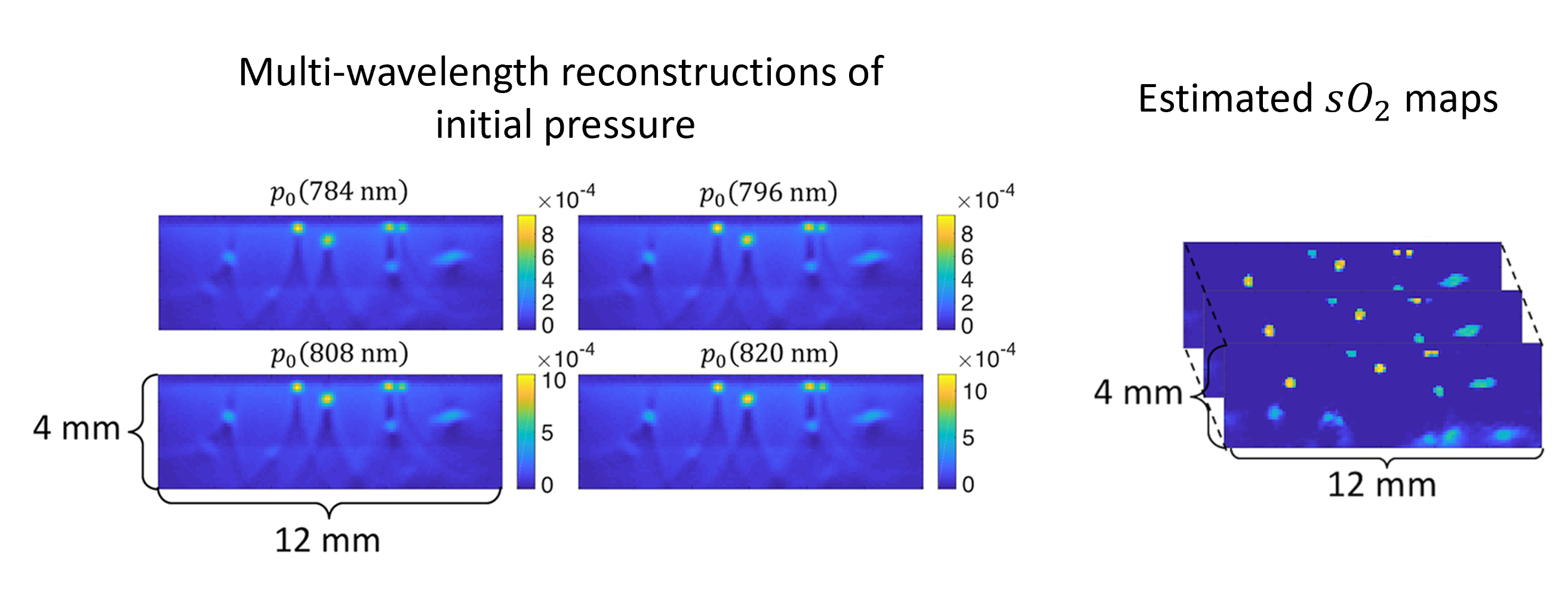}
\caption{\label{fig:oxygenation_estimation} Direct estimation of blood oxygenation saturation maps from multi-wavelength acoustic reconstructions of initial pressure in 3D of simulated human tissue phantoms.  (Left) Input slices (through 3D volume) of reconstructed initial pressure at 4 wavelengths. (Right) Neural network estimation of $sO_2$ maps in 3D. (Figure adapted from \cite{bench2020toward})}
\end{figure}

Similarly as for the acoustic problem, data-driven approaches have been utilised recently to solve the optical part, albeit there are fewer approaches proposed. A common approach follows the two-step method ii., where acoustic reconstructions are available. Then, a neural network is trained to recover the absorption coefficient distribution \cite{chen2020deep}. 
As mentioned before, there is often interest in computing related quantities such as the blood oxygen saturation $sO_2$. A few works have proposed to use a neural network similarly as in the previous approach, but instead of recovering the absorption coefficient, the network is trained to directly produce $sO_2$ maps from multi-wavelength acoustic reconstructions \cite{bench2020toward,grohl2021learned}, see Fig. \ref{fig:oxygenation_estimation} for an illustration.
Currently, there are no equivalent approaches to the model-based iterative paradigm iii. for the optical part, but there are studies for the related problem of DOT. Here, the authors proposed a learned iterative update for the Gauss-Newton algorithm \cite{mozumder2021model}.

\subsection{Single-stage approaches}

In addition to the two-step approach where acoustic and optical inverse problems are solved one after another, estimation of the optical parameters directly from the photoacoustic measurements have also been studied.
This one-step  approach was formulated utilising the Born approximation in \cite{shao2012iterative, song2014considering}. 
Furthermore, one-step approaches using Tikhonov regularisation \cite{haltmeier2015single},
$\ell_1$ sparsity regularisation  \cite{gao2015limited}, and total variation regularisation \cite{javaherian2019direct} have been considered. 
Further, a Bayesian approach  for the single-stage approach was formulated in \cite{pulkkinen2016direct}.
In \cite{ding2015one}, a one-step approach was used to recover the speed of sound and optical parameters simultaneously.

The authors in \cite{grohl2018confidence} investigated the possibility to directly recover optical parameters from the measured acoustic signal using machine learning methods. 
It was reported that such a naive approach is generally outperformed, especially in terms of generalisation capabilities to unseen data, by methods that use intermediate data or model information.

\section{Conclusions and outlook}
Much of the theory and analytical approaches on reconstructions for quantitative PAT assumes ideal measurement environments. 
In practice, however, the photoacoustic signal is affected by the measurement system, for example bandwidth and directivity of the acoustic sensors, as well as uncertainties related to the measurement setup.
Therefore, a complete modelling of photoacoustic data requires, in addition to modelling optical and acoustic phenomena, modelling of measurement system specifics. 
Furthermore, regardless of carefully considering all measurement setup specifics, modelling of them always contains uncertainties.
Examples of such uncertainties include, e.g., uncertainties in modelling the locations of the ultrasound sensors, variations in the light illumination, uncertainties in the frequency response of the ultrasound sensors, etc.
These all can affect the accuracy of the image reconstruction methodology and result into artefacts in images, errors in quantitative estimates, and less reliable confidence intervals.
Compensating errors and uncertainties in photoacoustics have been studied, for example, using Bayesian approximation error modelling \cite{hanninen2020application,sahlstrom2020modeling,tarvainen2013bayesian} and utilising learned model correction methods \cite{lunz2021learned}.

Additionally, fundamental challenges are given by the physics. One such  problem is given by the speed of sound that is often assumed to be constant and even when assumed constant, the accurate value inside the tissue may still not be known. 
In principle, a spatially-varying speed of sound could be recovered jointly with the acoustic reconstruction, although this problem is inherently unstable and requires additional data or prior knowledge \cite{matthews2018parameterized, kirsch2012simultaneous, liu2015determining, matthews2017joint}. 
One possibility here is the use of a separate modality, such as ultrasound tomography \cite{jose2012, xia2013enhancement, mercep2019}, to recover a speed of sound map prior to reconstruction. 
More generally, in the Bayesian setting, errors due to uncertainties in the speed of sound can be compensated for using Bayesian approximation error modelling \cite{tick2019modelling}, but the method is limited to small variations in the speed of sound or prior knowledge of the regions with larger variations. 
Here, data-driven methods can be helpful when the uncertainty on model parameters is incorporated into the training data. In this case, empirical evidence shows that the networks can learn to compensate for this uncertainty in the learned reconstruction.

Finally, the most accurate and stable reconstructions can still be obtained by model-based techniques, but the need to repeatedly evaluate the model equations leads to an immense computational overhead, that limits real-time applicability. This is even true for learned methods that do involve physics modelling to some extent, either in a two-step reconstruction or in learned iterative approaches. An area of active research considers how to effectively cut-down the computational bottle neck of the involved models by using model reduction techniques (coarser discretistations) or approximate models (making certain simplifying assumptions) \cite{cox2005,hauptmann2018approximate}. This results in a very similar problem as in model uncertainties, except that these are now introduced as additional complication. Here, the data-driven paradigm is especially promising, as we can effectively simulate training data and compensate the additional introduced approximation errors using neural networks \cite{hauptmann2023model,lunz2021learned,smyl2021learning}.

In conclusion, while the basic mathematical problem of quantitative PAT is solved, there are still many problems that need to be considered to achieve reliable and fast quantitative reconstructions in practice. The primary challenge can be identified as arising uncertainties when imaging \emph{in-vivo}, as opposed to ideal simulated studies, as well as computational limitations that one encounters in high-dimensional and possibly even temporal data. Data-driven approaches offer some new opportunities, but are certainly not the only solution, as their generalisation capabilities and robustness to changing imaging setups is still not fully understood. This emphasises the need for fundamental research to further our basic understanding, in combination with modern data-driven techniques.

\section*{Acknowledgements}
This work was supported by the European Research Council (ERC) under the European Union’s Horizon 2020 research and innovation programme (grant agreement no. 101001417 - QUANTOM) and the Academy of Finland (Centre of Excellence in Inverse Modelling and Imaging projects 353093 and 353086, and the Flagship Program Photonics Research and Innovation grant 320166, and Academy Research Fellow project 338408).
The authors would like to thank the Isaac Newton Institute for Mathematical Sciences, Cambridge, for support and hospitality during the programme Rich and Nonlinear Tomography where work on this paper was undertaken, supported by EPSRC grant no EP/R014604/1.

\bibliography{references}   
\bibliographystyle{siam}

\end{document}